\documentstyle[aps,preprint]{revtex}
\def\ltsim{\mathop{\raise3pt\hbox{$<$}\llap{\lower3pt\hbox{$\sim
$}}}}
\def\gtsim{\mathop{\raise3pt\hbox{$>$}\llap{\lower3pt\hbox{$\sim
$}}}}
\begin{document}
\tightenlines
\draft
\preprint{}
%\twocolumn
%[\hsize\textwidth\columnwidth\hsize\csname @twocolumnfalse\endcsname

\title{Quantum Monte Carlo simulations of infinitely strongly correlated 
       fermions}
\author{M.\ Brunner and  A.\ Muramatsu}
\address{Institut f\"ur Physik, Universit\"at Augsburg, 
Memmingerstra\ss e 6, D-86135 Augsburg,\\
and \\
$^\dagger$Institut f\"ur Theoretische Physik III, Universit\"at Stuttgart,
Pfaffenwaldring 57, D-70550 Stuttgart,\\ 
Federal Republic of Germany}
\date{\today}
\maketitle
\begin{abstract}
Numerical simulations of the two-dimensional t-J model in the limit 
$J/t \ll 1$ are performed for rather large systems (up to 
$N = 12 \times 12$) using a world-line loop-algorithm. It is shown that
in the one-hole case with $J=0$, where no minus signs appear, very low 
temperatures ($\beta t \sim 3000$) are necessary in order to reach Nagaoka's 
state. $J/t \ltsim 0.05$ leads to the formation of partially polarized 
systems, whereas $J/t \gtsim 0.05 $ corresponds to minimal 
spin. The two-hole case shows enhanced total spin up to the lowest attainable
temperatures ($\beta t = 150$).
\end{abstract}
%]

\narrowtext

\par
Since the discovery of high-T$_{\rm c}$ superconductors (HTS), a great deal 
of interest was focused on strongly correlated systems like the Hubbard and 
t-J models. Especially the latter system attracts a lot of attention since
it can be viewed as a model on its own right for HTS \cite{zhang88} and not 
only as the strong coupling limit of the Hubbard model. A key method for their 
understanding is provided by numerical simulations. For the Hubbard model, 
quantum Monte Carlo (QMC) simulations \cite{scalapino94} can be performed 
for not too large values of the interaction $U$ ($\ltsim 12 $ in units of 
the hopping matrix element $t$) and also not too large inverse temperatures 
$\beta$ ($ \ltsim 10/t$), except for the half-filled case, where the so-called 
minus-sign problem is absent, limiting thus our understanding of the doped 
phase at low temperatures in the strong coupling limit. On the other hand, 
exact diagonalizations of the t-J model \cite{dagotto94} provide information 
of precisely that limit but confined to rather small sizes $N$ ($\ltsim 30$). 
QMC simulations for the t-J model were successfully performed in one-dimension
\cite{assaad91}, whereas previous simulations in higher dimensions suffered
under metastability problems \cite{zhang91} or were restricted to the 
computation of energies in the ground state \cite{boninsegni92}. Further 
numerical studies were carried out using high-temperature expansions (HTE)
\cite{puti1,puti2} and density-matrix renormalization-group (DMRG)
\cite{white96,Liang95}.  
\par
In this Letter QMC simulations of the t-J model are presented for the first
time for large systems ($N \ltsim 12 \times 12$) and a large temperature 
range. 
Large system sizes are necessary especially in the case $J=0$, as shown 
by Barbieri et al.\ \cite{barbieri90}, where a change in the stability of the 
fully polarized ferromagnetic state is observed for two holes at around 
$N=12 \times 12$.
The simulations are based on a recently developed
world-line loop-algorithm \cite{kawashima94}. We use a representation of
the t-J model where it is obvious that it has no minus sign problem in the
case of one hole and $J=0$ \cite{khaliullin90,antimo95}. As shown below, 
this enables us to reach very low temperatures ($\beta t \sim 3000$), that are
necessary to converge to Nagaoka's ferromagnetic ground state. 
Such a slow convergence suggests that a high density of low lying 
excitations is present. We interpret them in the frame of a spin-wave theory, 
such that an effective ferromagnetic exchange interaction $J_{eff}$ can be 
assigned for each lattice size. Finite-size scaling shows, that $J_{eff}
\rightarrow 0$ in the thermodynamic limit as is expected. Hence, Nagaoka's 
ferromagnetic state has zero spin-stiffness in that limit. A frequently 
addressed question is the stability of Nagaoka's state for $J \neq 0$ 
or more than one hole \cite{doucot89,fang89,wurth95}. 
Our simulations show 
that for $N=10 \times 10$, $J\gtsim 5 \times 10^{-4}$ suffices to bring the 
system to a ground-state without maximal spin. 
Due to the minus-sign problem, 
simulations for more than one-hole could not be performed at temperatures
as low as in the one-hole case ($\beta t \ltsim 150$) 
and less conclusive 
results could be obtained.
However, it should be remarked that such temperatures are by far lower than
those reached previously by either QMC ($\beta t \ltsim 2$) \cite{zhang91}
or HTE ($\beta t \ltsim 5$) \cite{puti2}.
Finally, on increasing $J$, we show that the system goes to minimal 
spin at $J/t \gtsim 0.05$. At smaller $J$ the system is partially polarized.
\par
We first describe shortly the representation of the t-J model used in the 
simulations. Exact operator identities lead to a decomposition of the 
standard creation ($c_{i,s}^\dagger$) and annihilation ($c_{i,s}$)operators 
for fermions with spin $s = \uparrow, \downarrow$ \cite{khaliullin90,antimo95} 
\begin{eqnarray}
c^\dagger_{i,\uparrow} = \gamma_{i,+} f_i - \gamma_{i,-} f_i^\dagger \, , 
\; \; \;
c^\dagger_{i,\downarrow} = \sigma_{i,-} (f_i + f_i^\dagger) \, ,
\end{eqnarray}
where $\gamma_{i,\pm} = (1 \pm \sigma_{i,z})/2 \, , \; \sigma_{i,\pm} =
(\sigma_{i,x} \pm i \sigma_{i,y})/2$, the spinless fermion operators fulfill
the canonical anticommutation relations $\{f_i^\dagger,f_j\} = \delta_{ij}$,
and $\sigma_{i,\alpha}\, , \; \alpha = x,y,$ or $z$ are the Pauli matrices. 
The constraint to avoid doubly occupied states reduces in this case to 
$\sum_i \gamma_{i,-} f^\dagger_i f_i = 0$ which is a holonomic constraint 
in contrast to the one normally used with the standard representation. 
Moreover, this constraint commutes with the Hamiltonian (\ref{tj}), such that 
once the simulation is prepared in that subspace, it remains there in the 
course of the evolution in imaginary time. The t-J Hamiltonian in this new 
representation has the following form
\begin{eqnarray}
\label{tj}
H_{t-J} = t \sum_{<i,j>} \left( P_{ij} f^\dagger_i f_j 
        + \frac{J}{2} \Delta_{ij} (P_{ij} -1) \right) \; ,
\end{eqnarray}
where $P_{ij} = (1+\vec{\sigma}_i \cdot \vec{\sigma}_j)/2$, 
$\Delta_{ij} = (1 - n_i - n_j)$, and $n_i = f^\dagger_i f_i$. The constant 
added in (\ref{tj}) ensures that $H_{t-J}$ reduces to the standard t-J model.
In the case of one hole with $J=0$, there is only one fermion present and the 
simulation has no minus-sign problem due to the exchange of fermions. For 
$J>0$ a minus sign can occur because the exchange of two spins in space time
generates them.  
\par
Next we describe the world-line loop-algorithm. The world-lines are as 
usually defined by checkerboarding space-time \cite{hirsch82}. However, a 
direct application of this algorithm in two spatial dimensions is not 
effective in dealing with identical particles, since each permutation of any 
two particles represents a different sector that cannot be reached from 
another one. Especially in the case of fermions, the minus signs would lead 
to an extremely innacurate estimate of any observable. This difficulty can
be overcome with a loop algorithm \cite{kawashima94}, which does not preserve 
the linking topology of the world-lines. Since we are dealing with three states
per site, two world lines are introduced (one for the fermions and one for the
spin-ups) and their updating is performed separately, the sequence being 
chosen at random. Technical details will be published elsewhere \cite{us}. 
\par
In the following we describe the results for the limit $J \rightarrow 0$ and
mainly with one hole. Figure \ref{fig1hole} shows the structure form factor
\begin{eqnarray}
S({\vec q})= \frac{1}{N}
\sum\limits_{{\vec x}_i} \left(exp\left(\imath {\vec x}_i \cdot
\vec q\right) \sum_j S^z_j S^z_{j+{\vec x}_i}\right)
\end{eqnarray}
at $\vec{q}=0$, i.e.\ $<{S^{z}_{Tot}}^2>$ for $N=8 \times 8$ and 
$N=10 \times 10$ sites, as a function of $\beta t$. 
As can be clearly seen, very low temperatures are needed
in order to reach the fully saturated ferromagnet as predicted by Nagaoka's
theorem. This suggests that a high density of low energy excitations is 
present as would be expected in the case of a Heisenberg ferromagnet. As a 
simple test of this hypothesis, we compare the temperature dependence
of $S(\vec{q}=0)$ from the QMC simulation with a spin-wave result, where the
ferromagnetic exchange coupling $J_{eff}$ of the hypothetical Heisenberg 
ferromagnet is chosen such as to fit the results of the simulation in 
Fig.\ \ref{fig1hole}. Figure \ref{sw} shows that $J_{eff} \propto 1/N$, such
that in the thermodynamic limit the spin-stiffness vanishes.   
\par
Figure \ref{fig2hole} shows that, at the attainable temperatures, the 
ferromagnetic correlations are enhanced by adding a second hole to the system
($N = 12 \times 12$), suggesting a picture of a ferromagnetic polaron 
surrounding each hole, with a size limited by thermal fluctuations. Our 
result is consistent with a stability analysis by Barbieri et al.\ 
\cite{barbieri90}, where the Nagaoka state is stable for two holes and 
$N \gtsim 12 \times 12$. However, such a comparison should be taken with care 
since the results of Ref.\ \cite{barbieri90} are strictly valid in the limit
$N_h \ll \ln N$, where $N_h$ is the number of holes. Also DMRG studies 
\cite{Liang95} support 
a Nagaoka state for such a low doping ($\delta \sim 1.4\%$), although the 
geometry used in that case is quite different from ours. Unfortunately, the 
minus-sign problem precludes simulations at lower temperatures, such that 
the ground state cannot be reached. In fact, by comparing Fig.\ 
\ref{fig2hole} with Fig.\ \ref{fig1hole}, it is clear that in order to 
extract conclusions about the total spin in the ground-state, temperatures 
much lower than $\beta t = 150$ are necessary, such that previous assertions 
on the character of the ground state from results at $\beta t = 2$ to 5 
\cite{zhang91,puti2} are doubtful.
\par
Exact diagonalization studies have shown, that the Nagaoka state breaks down 
for $J > J_1 \sim 0.075t$ in a $4 \times 4$ system \cite{bonca89}.
Further they show, that the system goes to minimal spin at $J_2\gtrsim 0.088$.
Between $J_1$ and $J_2$ the system is partially polarized.
In the following we show, that for larger systems ($N = 10 \times 10$)
$J_1$ and $J_2$ are of different order of magnitude. Whereas Fig.\ 
\ref{S_0J_a} shows that the system is partially polarized for $J=0.0005 t$,
Fig.\ \ref{S_0J_c} demonstrates that for $J \gtsim 0.05 t$ the
system goes to a state with minimal total spin. At the same time, 
antiferromagnetic correlations begin to dominate at low temperatures. Although
the minus sign problem does not allow a very accurate determination, our data
show, that the system is partially polarized for 
$0.0005 \ltsim J \ltsim 0.05$. The last boundary is somewhat higher than
the lower bound proposed for the Hubbard-model to reach 
an antiferromagnetic state in the one-hole case 
($U^{AF}_c > (4t/\pi) N \ln N$) \cite{kusukabe94}. 

\par
Summarizing, we have performed QMC simulations of fermions in the limit of 
infinitely strong correlations. The world-line loop-algorithm allows 
simulations of rather large systems ($N \ltsim 12 \times 12$) and very
low temperatures (up to $\beta t \sim 3000$). Such temperatures are necessary 
in order to reach the fully polarized state in the case of one-hole and $J=0$. 
$J/t \ltsim 0.05$ leads to the formation of partially polarized 
systems, whereas $J/t \gtsim 0.05 $ corresponds to minimal 
spin. The two-hole case shows enhanced total spin up to the lowest attainable
temperatures ($\beta t = 150$).
  
We are very gratefull to Antimo Angelucci for instructive discussions on
the initial stages of the project. The calculations were performed at LRZ 
M\"unchen and HLR-Stuttgart. We thank the above institutions for their support.

$^\dagger$ Permanent address.
\bibliographystyle{./prsty}
\bibliography{setenza}
\onecolumn
\begin{figure}[H]
\caption{$S(\vec q=0)/N$ for  $N = 8 \times 8$ and $10 \times 10$,
with one hole at $J=0$. The dotted line
represents the fully saturated ground state. The full lines
correspond to $S(\vec q=0)$ of a Heisenberg ferromagnet in spin wave theory
with $J=0.0065 t$ and
$J=0.0030 t$, respectively.\\}\label{fig1hole}
\end{figure}

\begin{figure}
\caption{$J_{eff}$ of the Nagaoka state for different lattice sizes $N$.\\
\hspace{40cm}}\label{sw}
\end{figure}
\begin{figure}
\caption{Comparison of one hole and two holes in a $12 \times 12$ lattice. One sees, that
ferromagnetism is enhanced by adding another hole.\\}\label{fig2hole}
\end{figure}

\begin{figure}
\caption{$S(\vec q=0)$ for a $10 \times 10$ lattice with one hole at
$t\gg J>0$.\\\hspace{40cm}}\label{S_0J_a}
\end{figure}

\begin{figure}
\caption{$S(\vec q)$ for $\vec q=0$ and $\vec q=(\pi,\pi)$
for a $10 \times 10$ lattice with one hole at
$J=0.05 t$\\\hspace{40cm}}\label{S_0J_c}
\end{figure}

\newpage

\begin{figure}[t]
\input{Sq10SW.sty}
%\caption{$S(\vec q=0)/N$ for  $N = 8 \times 8$ and $10 \times 10$, 
%with one hole at $J=0$. The dotted line
%represents the fully saturated ground state. The full lines
%correspond to $S(\vec q=0)$ of a Heisenberg ferromagnet in spin wave theory
%with $J=0.0065 t$ and 
%$J=0.0030 t$, respectively.}\label{fig1hole}
\end{figure}
%
%\begin{figure}
%\input{Sq8SW.sty}
%\caption{$S(\vec q=)$ for a $\times 10$ lattice.
%with one hole at $J=0$. The dashed line
%represents the fully saturated ground state. The full line
%shows $S(\vec q=0)$ of a Heisenberg ferromagnet ($J=0.008t$)
%in spin wave theory.}\label{fig1hole8}
%\end{figure}
%
\newpage
\begin{figure}[t]
\input{Jeff.sty}
%\caption{$J_{eff}$ of the Nagaoka state for different lattice sizes $N$.}\label{sw}
\end{figure}
\newpage
\begin{figure}[t]
\input{Sq12n1n2.sty}
%\caption{Comparison of one hole and two holes in a $12 \times 12$ lattice. One sees, that
%ferromagnetism is enhanced by adding another hole.}\label{fig2hole}
\end{figure}
\newpage
\begin{figure}[t]
\input{Sq010Ja.sty}
%\caption{$S(\vec q=0)$ for a $10 \times 10$ lattice with one hole at
%$t\gg J>0$.}\label{S_0J_a}
\end{figure}
%
%\begin{figure}
%\input{Sq0piJa.sty}
%\caption{$S(\vec q)$ for $\vec q=0$ and $\vec q=(\pi,\pi)$
%for a $10 \times 10$ lattice with one hole at
%$J=0.1 t$}\label{S_0J_b}
%\end{figure}
%
\newpage
\begin{figure}[t]
\input{Sq0piJa05.sty}
%\caption{$S(\vec q)$ for $\vec q=0$ and $\vec q=(\pi,\pi)$
%for a $10 \times 10$ lattice with one hole at
%$J=0.05 t$}\label{S_0J_c}
\end{figure}

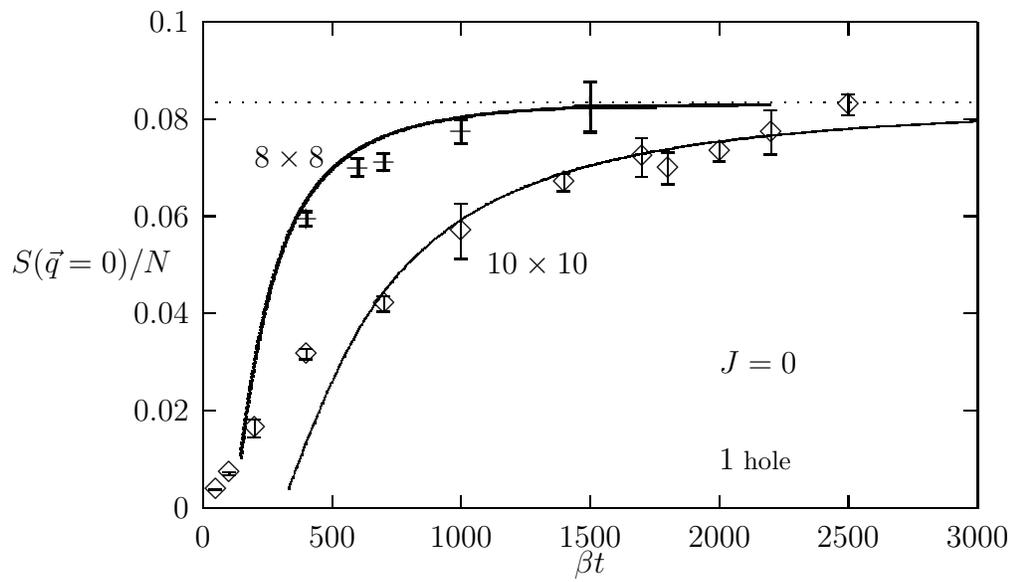
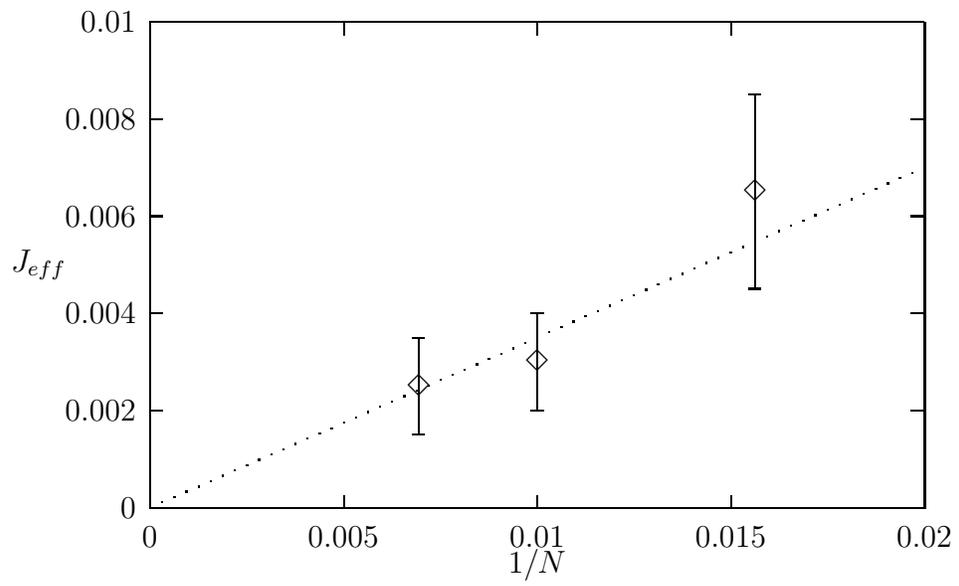
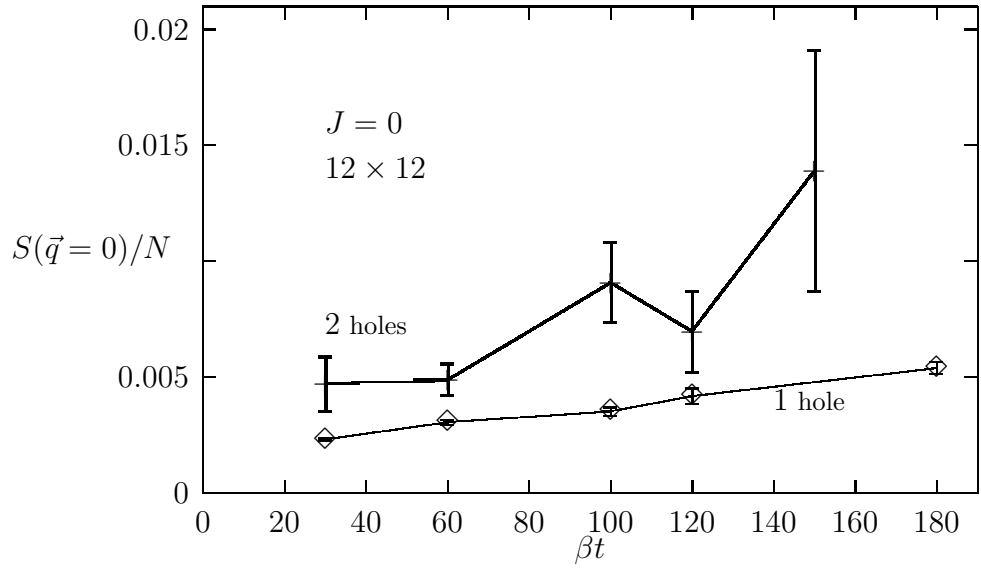
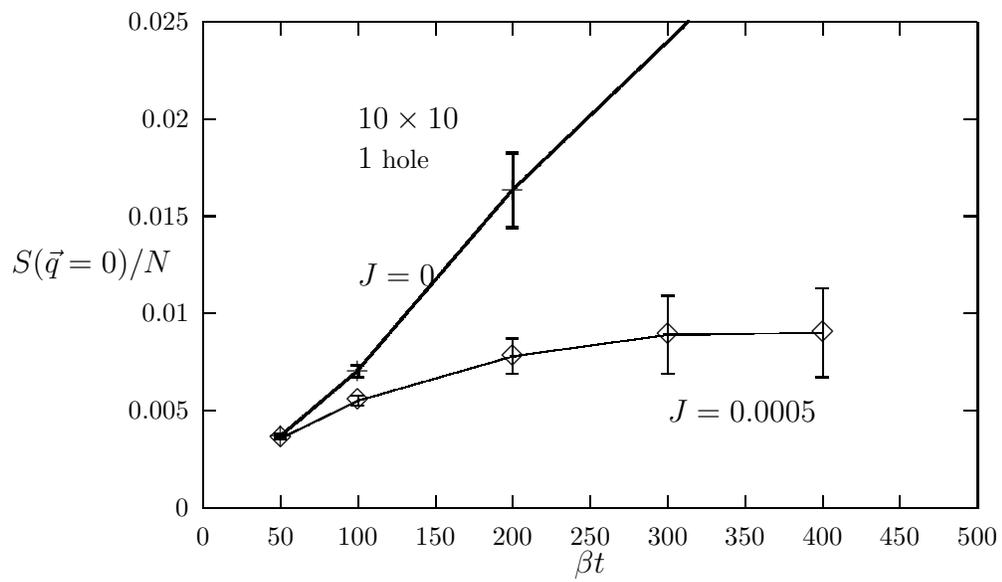
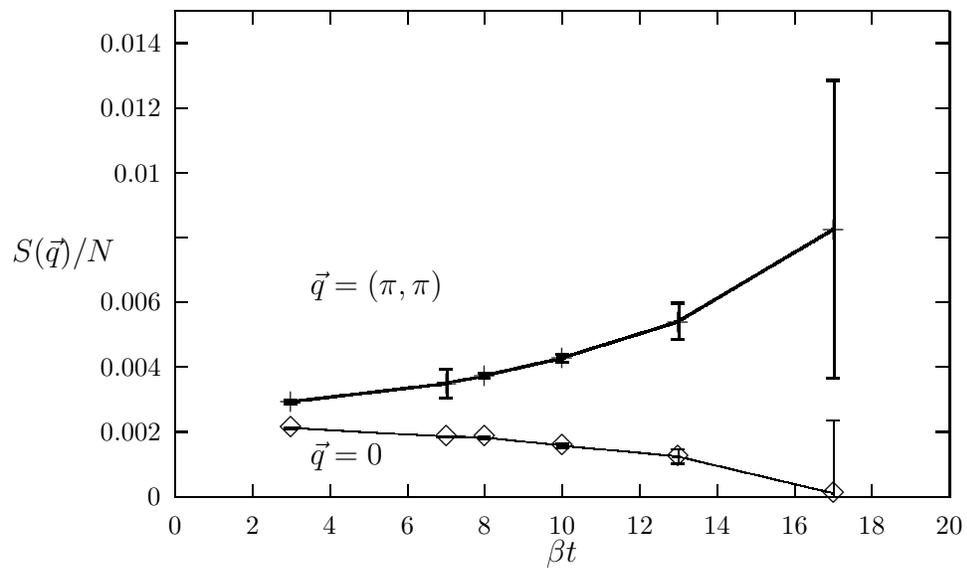
\end{document}